\documentclass[twocolumn,aps,prl,showpacs]{revtex4-1}

\usepackage{graphicx}
\usepackage{amsfonts}
\usepackage{amsmath}
\usepackage{amssymb}

\def\endproof{\vrule height6pt width6pt depth0pt}

\begin{document}


\title{Proposal for revealing quantum nonlocality via local contextuality}


\author{Ad\'an Cabello}
\email{adan@us.es}
\affiliation{Departamento de F\'{\i}sica Aplicada II, Universidad de
Sevilla, E-41012 Sevilla, Spain}


\date{\today}



\begin{abstract}
Two distant systems can exhibit quantum nonlocality even though
the correlations between them admit a local model. This
nonlocality can be revealed by testing extra correlations
between successive measurements on one of the systems which do
not admit a noncontextual model whatever the reduced state of
this system is. This shows that quantum contextuality plays a
fundamental role in quantum nonlocality, and allows an
experimental test of the Kochen-Specker with locality theorem.
\end{abstract}



\maketitle


{\em Introduction.---}One of the most striking aspects of
quantum mechanics (QM) is quantum nonlocality; that is, the
impossibility of reproducing quantum correlations in terms of
classical local hidden variable (HV) theories \cite{Bell64}. In
this Letter we will show that there is a fundamentally
different way to reveal quantum nonlocality which does not
involve classically inexplicable correlations between distant
systems. For this purpose, we derive a Bell inequality which is
violated by QM. The important point is that the quantum
violation occurs even though the correlations between the main
system and the auxiliary one can be reproduced with a local
model. The obstacle for classical local theories is the
state-independent contextuality of one system \cite{Cabello08,
BBCP09} rather than the correlations between distant systems.
This shows that quantum contextuality, a property of quantum
systems with more than two states, plays a fundamental role in
quantum nonlocality, and provides a different way to
experimentally observe quantum nonlocality, valid for any
system with $d>2$ states entangled with an auxiliary system
with $d$ states. We will illustrate it with an inequality
violated if $d \ge 4$, which is particularly simple, but the
method can be also applied to any system with $d>2$.

This inequality serves some additional purposes. The first is
to point out that, contrary to a common belief (see, e.g.,
\cite{GTTZ09}), there is something new to learn about quantum
nonlocality from the proofs of the so-called Kochen-Specker
(KS) with locality theorem of impossibility of local HV
theories \cite{Kochen, HR83, Stairs83, Redhead87, BS90}, or
``free will'' theorem \cite{CK06, CK09}, which cannot be
learned from other proofs of quantum nonlocality like
\cite{Bell64, CHSH69, GHZ89, GHSZ90, Mermin90b}. The second is
to elude the criticisms to some recent experiments to test the
KS theorem of impossibility of noncontextual HV theories
\cite{Specker60, Bell66, KS67} with ions \cite{KZGKGCBR09},
neutrons \cite{BKSSCRH09}, photons \cite{ARBC09, LHGSZLG09},
and nuclear magnetic resonance systems \cite{MRCL09}. The
problem of noncontextual HV theories is that the assumption of
noncontextuality is not motivated by a physical principle, like
locality in Bell inequalities, so one might think that
noncontextual theories are physically unplausible \cite{Bell66,
Bell82}, since there are classical models reproducing the
results of these experiments \cite{La Cour09a, GKCLKZGR10}. A
third purpose is to avoid a loophole in these experiments due
to the nonperfect compatibility of sequential measurements
\cite{KZGKGCBR09, GKCLKZGR10}.


\begin{figure}[tb]
\center
\includegraphics[width=1.00\linewidth]{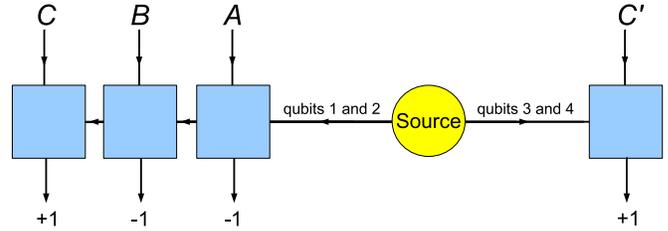}
\caption{Scheme of the experiment to reveal quantum nonlocality by local contextuality.
Alice performs three compatible
dichotomic measurements sequentially, for example, $A$, $B$, and $C$
on her system (qubits 1 and 2), while Bob performs a single measurement,
for example, $C'$ on his system (qubits 3 and 4). According to QM, the product of
the results of $A$, $B$, and $C$ can be predicted with certainty whatever the state of Alice's system.
The experiment also tests the correlations between Alice and Bob's systems.
According to QM, in the state \eqref{state}, the results
of $C$ and $C'$ are perfectly correlated.} \label{setup}
\end{figure}


{\em Scenario.---}Consider four qubits distributed between two
distant locations. Alice has qubits 1 and 2, and Bob has qubits
3 and 4. In each run of the experiment, Alice performs three
successive compatible measurements on the subsystem composed of
qubits 1 and 2, and Bob performs a single measurement on the
subsystem composed of qubits 3 and 4; as illustrated in Fig.
\ref{setup}. The separation between Alice and Bob's
measurements inhibits any communication between Alice's (Bob's)
choices of measurements and Bob's (Alice's) results. The four
qubits are initially prepared in the state
\begin{equation}
|\Psi \rangle_{1234} = |\psi^-\rangle_{13} \otimes
|\psi^-\rangle_{24}, \label{state}
\end{equation}
where $|\psi^-\rangle_{ij} = \frac{1}{\sqrt{2}} (|0\rangle_i
\otimes |1\rangle_j-|1\rangle_i \otimes |0\rangle_j)$. On
qubits 1 and 2, Alice sequentially measures one of the six
sequences: $ABC$ (i.e., first she measures $A$, then $B$, and
finally $C$; see Fig. \ref{setup}), $bac$, $\gamma \beta
\alpha$, $A a \alpha$, $b B \beta$, or $\gamma c C$, where:
\begin{align}
&A=z_{1},\;\;\;\;\;\;\;\;\;\;\;\;B=z_{2},&C=z_{1} z_{2}, \nonumber \\
&a=x_{2},\;\;\;\;\;\;\;\;\;\;\;\;b=x_{1},&c=x_{1} x_{2}, \nonumber \\
&\alpha=z_{1} x_{2},\;\;\;\;\;\;\;\;\;\beta=x_{1} z_{2},&\gamma=y_{1} y_{2},
\label{observables}
\end{align}
and $z_{1} x_{2}=\sigma_z^{(1)} \otimes \sigma_x^{(2)}$, that
is, the tensor product of the Pauli matrices $Z$ of qubit 1 and
$X$ of qubit 2. Sequential measurements of this type have been
recently made on ions \cite{KZGKGCBR09} and photons
\cite{ARBC09}.

In the state \eqref{state}, each of the nine observables
\eqref{observables} of qubits 1 and 2 is perfectly correlated
or anticorrelated with the corresponding observable of qubits 3
and 4. In particular,
\begin{align}
 &\langle B B' \rangle = -1,\;\;\;\;\;\;
 \langle C C' \rangle = 1,\;\;\;\;\;\;\;\;\;
 \langle a a' \rangle = -1,\nonumber \\
 &\langle c c' \rangle = 1,\;\;\;\;\;\;\;\;\;\;\;\;
 \langle \alpha \alpha' \rangle = 1,\;\;\;\;\;\;\;\;\;\;
 \langle \beta \beta' \rangle = 1,
 \label{means}
\end{align}
where
\begin{align}
&B'=z_{4},\;\;\;\;\;\;\;\;\;\;\;\;C'=z_{3} z_{4},&a'=x_{4}, \nonumber \\
&c'=x_{3} x_{4},\;\;\;\;\;\;\;\;\;\;\alpha'=z_{3} x_{4},&\beta'=x_{3} z_{4}.
\label{Bobobservables}
\end{align}
Therefore, in the state \eqref{state}, the results of
$B,C,\ldots, \beta$ can be predicted with certainty from the
results of $B',C',\ldots, \beta'$, respectively. This
prediction is the same regardless of whether $\beta$ is
measured in the sequence $\gamma \beta \alpha$ or in the
sequence $b B \beta$.


{\em Bell inequality.---}Any local HV theory satisfies
\begin{equation}
 \langle \omega \rangle \equiv
 \langle \chi \rangle + \langle S \rangle \le 16,
\label{completeinequality}
\end{equation}
where
\begin{equation}
 \langle \chi \rangle \equiv
 \langle A B C \rangle
 + \langle b a c \rangle
 + \langle \gamma \beta \alpha \rangle
 + \langle A a \alpha \rangle
 + \langle b B \beta \rangle
 - \langle \gamma c C \rangle,
 \label{second}
\end{equation}
and $\langle A B C \rangle$ denotes the average of the product
of the outcomes of $A$, $B$, and $C$ measured in the sequence
$ABC$, and
\begin{equation}
 \begin{split}
 \langle S \rangle \equiv
  & |\langle B B' \rangle_{ABC}|
  + |\langle B B' \rangle_{bB\beta}|
  + |\langle C C' \rangle_{ABC}|
  + |\langle C C' \rangle_{\gamma c C}|
  \\
  & + |\langle a a' \rangle_{b a c}|
  + |\langle a a' \rangle_{A a \alpha}|
  + |\langle c c' \rangle_{bac}|
  + |\langle c c' \rangle_{\gamma c C}|
  \\
  & + |\langle \alpha \alpha' \rangle_{\gamma \beta \alpha}|
  + |\langle \alpha \alpha' \rangle_{A a \alpha}|
  + |\langle \beta \beta' \rangle_{\gamma \beta \alpha}|
  + |\langle \beta \beta' \rangle_{b B \beta}|,
 \label{beta}
 \end{split}
\end{equation}
where $\langle BB'\rangle_{ABC}$ denotes the average $\langle
BB' \rangle$ in those events where $B$ is measured in the
sequence $ABC$ on qubits 1 and 2, and $B'$ is measured alone on
qubits 3 and 4.


{\em Proof: }Let us denote by $\hat{B'}$ the outcome ($-1$ or
$1$) the local HV theory assigns to $B'$ {\em when no other
observable is measured before} $B'$. Similarly, $\hat{C'}$ is
the outcome the local HV theory assigns to $C'$ when no other
observable is measured before $C'$. In any local HV theory,
$\hat{B'}$ and $\hat{C'}$ are well defined even though both
cannot be simultaneously measured. Therefore, any local HV
theory must satisfy the following inequality:
\begin{equation}
 \langle A \hat{B'} \hat{C'} \rangle
 + \langle b \hat{a'} \hat{c'} \rangle
 + \langle \gamma \hat{\beta'} \hat{\alpha'} \rangle
 + \langle A \hat{a'} \hat{\alpha'} \rangle
 + \langle b \hat{B'} \hat{\beta'} \rangle
 - \langle \gamma \hat{c'} \hat{C'} \rangle
 \le 4,
 \label{secondbis}
\end{equation}
where the upper bound 4 can be obtained by checking all
possible combinations of outcomes ($-1$ or $+1$) for $A,
\hat{B'},\ldots, \gamma$.

Inequality \eqref{secondbis} is not directly testable because
$\hat{B'}$ and $\hat{C'}$ (or $\hat{a'}$ and $\hat{c'}$) cannot
be measured both in the first place. However, the following
sequence of inequalities:
\begin{equation}
  \begin{split}
    |\langle A & \hat{B'} \hat{C'} \rangle - \langle ABC \rangle|
  \\  & \le |\langle A \hat{B'} \hat{C'}\rangle - \langle A B \hat{C'}\rangle|
    +|\langle A B \hat{C'}\rangle -\langle ABC \rangle|
  \\ & \le \langle |A \hat{B'} \hat{C'} - A B \hat{C'} \hat{B'}^2|\rangle
   +\langle |A B \hat{C'} - A B C \hat{C'}^2|\rangle
  \\ & = \langle |A \hat{B'} \hat{C'} (1- B \hat{B'})|\rangle
   +\langle |A B \hat{C'} (1 - C \hat{C'})|\rangle
  \\ & \le 1 - |\langle BB' \rangle| + 1 - |\langle CC' \rangle|
  \label{eq:2}
  \end{split}
\end{equation}
allows us to see that the term $\langle A \hat{B'} \hat{C'}
\rangle$ in \eqref{secondbis} is lower bounded by
experimentally testable quantities,
\begin{equation}
\langle A \hat{B'} \hat{C'} \rangle \ge \langle ABC \rangle + |\langle BB' \rangle_{ABC}| + |\langle CC' \rangle_{ABC}|-2.
\end{equation}
Similarly, we can obtain experimentally testable lower bounds
for $\langle b \hat{a'} \hat{c'} \rangle$, $\langle \gamma
\hat{\beta'} \hat{\alpha'} \rangle$, $\langle A \hat{a'}
\hat{\alpha'} \rangle$, $\langle b \hat{B'} \hat{\beta'}
\rangle$, and $-\langle \gamma \hat{c'} \hat{C'} \rangle$.
Introducing all of them in \eqref{secondbis}, we obtain
inequality \eqref{completeinequality}.\hfill \endproof


{\em Quantum violation.---}The quantum prediction for the state
\eqref{state} is
\begin{equation}
 \langle \omega \rangle_{\rm QM} = 18,
\end{equation}
which violates inequality \eqref{completeinequality}. This
violation is due to a different reason than the violation of
previous Bell inequalities \cite{Bell64, CHSH69, Mermin90b}.
Inequality \eqref{completeinequality} has two terms: The term
$\langle S \rangle$, defined in \eqref{beta}, contains
correlations between the two distant systems. In the state
\eqref{state}, all of these correlations are trivial [see Eq.
\eqref{means}], thus $\langle S \rangle$ takes its maximum
value $\langle S \rangle=12$, and inequality
\eqref{completeinequality} becomes
\begin{equation}
 \langle \chi \rangle \le 4.
 \label{old}
\end{equation}
The term $\langle \chi \rangle$, defined in \eqref{second},
only contains correlations between successive measurements {\em
on one of the local systems} (Alice's). Inequality \eqref{old}
is similar to the inequality in Refs. \cite{Cabello08,
KZGKGCBR09, ARBC09, MRCL09} for noncontextual theories.
However, while in Refs. \cite{Cabello08, KZGKGCBR09, ARBC09,
MRCL09}, one assumed that if $A$ and $B$ are compatible, then
the outcome of $B$ does not depend on whether $A$ has been
measured before $B$, inequality \eqref{completeinequality}
holds for any local HV theory, even for those in which a
previous measurement of $A$ can change the outcome of $B$.

The difference between the maximum quantum violation and the
classical bound is the same for the inequality
\eqref{completeinequality} and the inequality \eqref{old}. In
both cases, the quantum violation occurs because, in QM,
\begin{equation}
\langle \chi \rangle_{\rm QM} = 6,
\end{equation}
since the product of the three operators representing $A$, $B$,
and $C$ is the identity $\openone$, and the same for $bac$,
$\gamma\beta\alpha$, $Aa\alpha$, and $bB\beta$, while it is
$-\openone$ for $\gamma cC$. This quantum violation is
independent of the state of Alice's system: It holds not only
when it is in a maximally mixed state [which is the reduced
state of qubits 1 and 2 when the state of the four qubits is
\eqref{state}], but also in any other state. Therefore, neither
entanglement nor the reduced state of Alice's system play a
role in the violation of inequality \eqref{old}. Consequently,
the role of entanglement in the violation of inequality
\eqref{completeinequality} is marginal. The role of the
entanglement of the state \eqref{state} is to allow us to
convert a test of state-independent quantum contextuality like
those in \cite{Cabello08, KZGKGCBR09, ARBC09, MRCL09} into a
Bell test. The only explanation in terms of HV of the results
observed in previous experiments \cite{KZGKGCBR09, ARBC09,
MRCL09} is that the outcomes of the measurements on Alice's
system have changed due to previous measurements on Alice's
system, as in the models proposed in \cite{La Cour09a,
GKCLKZGR10}. In the experiment proposed in this Letter, we can
test whether this is actually happening by testing wether the
expected perfect correlations Bob's system have changed due to
these previous measurements. We assume that the outcomes of
$A$, $b$, and $\gamma$ cannot change neither due to previous
local measurements (because they are always measured first),
nor due to spacelike separated measurements (because we assume
locality). However, in the sequence $ABC$, the outcomes of $B$
and $C$ could change by previous local measurements. To detect
a contextual behavior in the sequence $ABC$, we measure $ABCB'$
(i.e., we measure the sequence $ABC$ on Alice's system, and
$B'$ on Bob's) in half of the cases, and $ABCC'$ in the other
half, and then calculate $\langle BB'\rangle_{ABC}$ and
$\langle CC'\rangle_{ABC}$.

The entanglement of the state \eqref{state} can be replaced by
a different kind of entanglement. For example, Alice's
four-state system could belong to a four-system four-level
singlet state \cite{Cabello02}, and a violation of a Bell
inequality would occur due to the violation of \eqref{old} by
performing the same sequential measurements on Alice's system.


{\em Experimental requirements.---}As in any Bell experiment,
spacelike separation between one observer's choice of
measurements and the other observer's result is required if the
``upper bound to the speed with which information can be
effectively transmitted'' is assumed to be the speed of light
\cite{CK06}. We think that there is no fundamental obstacle to
satisfy this requirement (see, e.g., \cite{RWVHKCZW09}), but
even an experimental test of inequality
\eqref{completeinequality} without spacelike separation would
be important: If the results agree with QM, it would
demonstrate the impossibility of contextual explanations of the
results on Alice's system under the assumption that no mutual
disturbance occurs between the measurements on Alice's system,
and the measurements on Bob's, which is a more plausible
assumption than the unrestricted noncontextuality assumed in
previous experiments \cite{KZGKGCBR09, BKSSCRH09, ARBC09,
LHGSZLG09, MRCL09}. The question is whether such an experiment
is feasible.

A simple calculation shows that inequality
\eqref{completeinequality} is violated as long as
\begin{equation}
V > \frac{1}{4} \left(\sqrt{33-2 \langle \chi \rangle_{\rm{expt}}}-1
\right), \label{vis}
\end{equation}
where $V$ is the visibility of the prepared states, assuming
that, instead of a perfect singlet $|\psi^-\rangle$ in
\eqref{state}, we have $\rho = V |\psi^-\rangle \langle
\psi^-|+(1-V) \openone/4$, and $\langle \chi
\rangle_{\rm{expt}}$ is the observed value for $\langle \chi
\rangle$. The achieved visibility of two-qubit maximally
entangled states with two separated ions is $0.97$, which
corresponds to a fidelity $F=0.99$ \cite{BKRB08} (assuming that
$F=\frac{1}{2}\sqrt{3V+1}$). Then, according to \eqref{vis}, it
would be enough to observe $\langle \chi \rangle_{\rm{expt}}
> 4.59$ to violate \eqref{completeinequality}. The achieved violation of the
inequality \eqref{second} for a two-qubit maximally mixed state
is $5.30$ \cite{KZGKGCBR09}. For $\langle \chi
\rangle_{\rm{expt}} = 5.30$, inequality
\eqref{completeinequality} would be violated if $V > 0.93$.
These results suggest that it is feasible to observe a
violation of the inequality \eqref{completeinequality} with
trapped entangled ions. Similar promising results can be
obtained by combining the experimental fidelities of pairs of
hyperentangled photons (see, e.g., \cite{BDMVC06}) and the
observed violation of the inequality \eqref{second} for single
photons \cite{ARBC09}.


{\em Conclusions.---}We have shown that there is a different
way to experimentally reveal quantum nonlocality based on a
violation of a Bell inequality which involves not only
correlations between distant systems but also correlations
between successive measurements on one of the local systems.
The former can be reproduced with local models, the latter
cannot be reproduced using noncontextual models. The thing that
makes any model that tries to reproduce the trivial
correlations with an auxiliary distant system nonlocal is the
need of a contextual explanation of the state-independent
behavior of one of the systems. This result suggests that the
key quantum property behind quantum nonlocality is quantum
contextuality, a property of any quantum system with more than
two states, rather than only entanglement.

The inequality also provides an experimentally testable version
of the KS with locality or ``free will'' theorem \cite{Kochen,
HR83, Stairs83, Redhead87, BS90, CK06, CK09}, allows us to
experimentally exclude models proposed \cite{La Cour09a,
GKCLKZGR10} to explain the results of some recent experiments
\cite{KZGKGCBR09, ARBC09, MRCL09, BKSSCRH09, LHGSZLG09}, and
avoids the extra assumptions required when dealing with a
loophole specific to these experiments \cite{KZGKGCBR09,
GKCLKZGR10}.

An experimental violation of this inequality would not only
prove the impossibility of local theories and a class of
nonlocal theories (those in which the correlations between
successive measurements on one system are noncontextual), but
would shed new light on the origin of quantum nonlocality,
highlighting the importance of quantum contextuality.


\begin{acknowledgments}
The author thanks M.\ Barbieri, R.\ Blatt, M.\ Bourennane, D.\
P.\ DiVincenzo, Y.\ Hasegawa, M.\ Kleinmann, S.\ Kochen, B.\ La
Cour, H.\ Rauch, C.\ F.\ Roos, B.\ Terhal, H.\ Weinfurter, and
A.\ Zelinger for discussions, O. G{\"u}hne for pointing out a
hidden assumption in a previous version of the manuscript,
J.-\AA.\ Larsson, for improving the proof of inequality
\eqref{completeinequality}, and R.\ Laflamme for his
hospitality at the Institute for Quantum Computing, and
acknowledges support from the Spanish MCI Project No.\
FIS2008-05596, and the Junta de Andaluc\'{\i}a Excellence
Project No.\ P06-FQM-02243.
\end{acknowledgments}


\end{document}